# How a Randall-Sundrum Brane-World Effective Potential Influences Inflation Physics

A.W. Beckwith

*APS/ Contractor, Fermi National Laboratory, Menlo Park, CA USA 94025.*

*650-322-6768, abeckwith@UH.edu*

**Abstract.** In string theory, even when there are ten to the thousand power vacuum states, does inflation produce overwhelmingly one preferred type of vacuum state? We respond affirmatively to questions whether existence of graviton production is confirmable using present detector methodology. We use an explicit Randall-Sundrum brane-world effective potential as congruent with an inflationary quadratic potential start. This occurs after Bogomolnyi inequality eliminates need for ad hoc assumption of axion wall mass high temperature related disappearing. Graviton production has explicit links with a five-dimensional brane-world negative cosmological constant and a four-dimensional positive valued cosmological constant, whose temperature dependence permits an early universe graviton production activity burst. We show how di quarks, wave functions, and various forms tie into the Wheeler-De Witt equation. This permits investigating a discretized quantum bounce and a possible link to the initial phases of present universe's evolution with a prior universe's collapse to the bounce point—the initial starting point to inflationary expansion. This opens a possibility of realistically investigating gravitons as part of a space propulsion system and dealing with problems from a beam of gravity waves, which would create a g-force because the geodestic structure is near field. It can be applied to existing and to new space propulsion concepts.

## INTRODUCTION

To consider spin-two graviton detection in early universe conditions and brane-world generation of a large number of gravitons, we answer, affirmatively, Freeman Dyson's challenge to prove that it is feasible to ascertain criteria for the existence of gravitons where their likelihood of detection is so low in most astrophysical processes (Rothman and Boughn, 2006). Further, we present a different paradigm about the origins of dark energy—from the stand point of di quarks and an application of the Bogomolnyi inequality—so we can tie a model of baryogenesis into a collapse of an originally axion 'wall' dominated potential system to the Guth (1981) inflation quadratic potential. This relies upon an analytical connection (Park, Kim, and Tamarayan, 2006) between a four-dimensional cosmological 'constant' that we would expect from more normal 3+1 standard space-time metrics, to the admittedly quite unusual negative five-dimensional cosmological constant present in Randall-Sundrum brane-world models.

Data suggesting that the four-dimensional version of the "cosmological constant" in fact varies with background temperature. If temperature varies significantly during early universe baryogenesis, there would be a huge release of spin-two gravitons in early cosmic nucleation of a new universe. To answer whether "Even if there are $10^{1000}$ vacuum states produced by string theory, then does inflation produces overwhelmingly *one* preferred type of vacuum states over the other possible types of vacuum states (Guth, 2006)?" leads to a way to calculate the order of the electroweak phase transition, which may indeed fix a preferred vacuum state. This permits comparatively straightforward data collection of dark energy traces in the early universe by the JDEM satellite system, among other considerations. This is consistent with a flat space overlap of four- and five-dimensional Einstein-Hilbert least action integrals, which are the same value in 'empty space' but which diverge in magnitude (Loup, 2006) in non-empty energy-matter metric space time conditions. We begin by explicit calculations (Krauss, and Trodden, 1999) of the electroweak phase transition order via:

$$\frac{\left(\langle\phi\rangle\big|_{T=T_C} \equiv v(T_C)\right)}{T_C} \geq 1. \tag{1}$$

To incorporate more accurate reading of phase evolution and the minimum requirements of phase evolution behavior in a potential system permitting baryogenesis, imply using a Sundrum fifth-dimension. The fifth-dimension of the Randall-Sundrum brane-world (Sundrum, 2005) is, for $-\pi \leq \theta \leq \pi$, a circle map which is written, with $R$ as the radius of the compact dimension $x_5$ Circle maps were first proposed to model a free-spinning wheel weakly coupled by a spring to a motor. Circle map equations also describe a simplified model of the phase-locked loop in electronics. We use a circle map simply for a compact higher dimensional structure geometry, which is important in early universe geometries. A closed string can wind around a periodic dimension an integral number of times. Similar to the Kaluza-Klein case they contribute a momentum, as p = w R (w=0, 1, 2,...). The crucial difference is that this reverses with respect to the radius of the compact dimension, R. As the compact dimension becomes very small, these winding modes become very light! We write our fifth dimension (Sundrum, 2005) as:

$$x_5 \equiv R \cdot \theta. \tag{2}$$

This fifth dimension $x_5$ also creates an embedding potential structure leading to a complimentary embedded in five dimensions scalar field we model (Sundrum, 2005) as:

$$\phi(x^\mu, \theta) = \frac{1}{\sqrt{2 \cdot \pi \cdot R}} \cdot \left\{\phi_0(x) + \sum_{n=1}^{\infty} [\phi_n(x) \cdot \exp(i \cdot n \cdot \theta) + C.C.]\right\}. \tag{3}$$

Our embedded in four dimensions potential structure is modeled via the following phase transition listed in Eqn. (4). We refer to a scalar field using four-dimensional space-time, which we call $\tilde{\phi}$, and which leads to the potential having the following phase transformation (Beckwith, 2006a):

$$\begin{aligned}
\tilde{V}_1 &\to \tilde{V}_2 \\
\tilde{\phi}(increase) \leq 2 \cdot \pi &\to \tilde{\phi}(decrease) \leq 2 \cdot \pi \cdot \\
t \leq t_P &\to t \geq t_P + \delta \cdot t
\end{aligned} \tag{4}$$

Note that potentials $\tilde{V}_1$, and $\tilde{V}_2$ are two cosmological inflation potentials and that $t_p$, Planck time, is the time it would take a photon traveling at the speed of light to cross a distance equal to the Planck length, $\approx 5.39121(40) \times 10^{-44}$ seconds. Planck length, denoted by $l_P$, approximately 1.6 × $10^{-35}$ meters, is deemed "natural" because it can be defined from three fundamental physical constants: the speed of light, Plank's constant, and the gravitational constant. There is a phase transition between the first and second potentials, with a rising and falling value of the magnitude of the four dimensional scalar fields. When the scalar field rises corresponds to quantum nucleation of a vacuum state, represented by $\tilde{\phi}$, as we address later, there is a question whether there is a generic vacuum state as a starting point for the transformation to standard inflation given by the second scalar potential system. We describe the potentials $\tilde{V}_1$ and $\tilde{V}_2$ in terms of S-S' (soliton-anti soliton) style di quark pairs (Beckwith, 2006a) nucleating:

$$\tilde{V}_1(\phi) = \frac{M_P^2}{2} \cdot (1 - \cos(\tilde{\phi})) + \frac{m^2}{2} \cdot (\tilde{\phi} - \phi^*)^2. \tag{5}$$

$$\tilde{V}_2(\phi) \propto \frac{1}{2} \cdot (\tilde{\phi} - \phi_C)^2. \tag{6}$$

$\phi_C$ in Eqn. (6) is an equilibrium value of a true vacuum minimum after quantum tunneling through a barrier. Note that $M_P$ ($M_P$ = Planck's mass ≈ 1.2209 × $10^{19}$ GeV/$c^2$ = 2.176 × $10^{-8}$ kg) is the mass in which the Schwarzschild radius is equal to the Compton length divided by π; a Schwarzshild radius is proportional to the mass, with a proportionality constant involving the gravitational constant and the speed of light. The Schwarzschild radius formula can be found by setting the escape velocity to the speed of light; furthermore, where mass $m \ll M_P$, $m$ is the mass of the gravitating object. And as a final note, we have that a soliton is a self-reinforcing solitary wave caused by a delicate balance between nonlinear and dispersive effects in the medium. Solitons are found in many physical phenomena, as they arise as the solutions of a widespread class of weakly nonlinear dispersive partial differential equations describing physical systems. Here, we can visualize solitons as wave forms (Beckwith, 2006b) with the following properties:

1. They represent waves of permanent form;
2. They are localized, i.e., decay or approach a constant at infinity;
3. They can interact with other solitons but emerge from collision unchanged apart from a phase shift.

We assign to a soliton-anti soliton pair (S-S') equivalence to di quarks. As a convention, we define constituent masses for di quarks as the sum of the respective quark masses, while di quarks are a combination of quarks in the color antitriplet $3 \otimes 3 = \overline{3} \oplus 6$. Their most attractive linkage as to new inflation physics models is as cooper pairs of color superconductivity, which permits simplification of SUSY model physics. Briefly, we assign a bosonic counter part to each fermion, which can be seen in the assignment of a counter part to a quark, a so called squark. This procedure is greatly aided in the di quark model, which simplifies matters immensely.

The overall transformation from Eqn. (5) to Eqn (6) is outlined in the (Coleman, 1977) paper on false vacuum nucleation. We note that $\phi^*$ in Eqn. (6) is a measure of the onset of quantum fluctuations where fluctuations have an upper bound of $\tilde{\phi}$ (here, $\tilde{\phi} > \phi_C$) and $\tilde{\phi} \equiv \tilde{\tilde{\phi}} - mt/\sqrt{12 \cdot \pi \cdot G}$, where we use $\tilde{\tilde{\phi}} > \sqrt{60/2\pi} M_P \approx 3.1 M_P \equiv 3.1$ (Guth, 2000).

## RANDALL SUNDRUM EFFECTIVE POTENTIAL

The consequences of the fifth-dimension, considered in Eqn. (5), show up in a simple warped compactification involving two branes: a Planck world brane, and an IR brane (Sundrum, 2005). The Planck brane; the brane where gravity is localized, is a four-dimensional structure defining the standard model universe. The IR brane is used as structure to facilitate solving five-dimensional Einstein equations. In quantum field theory physics, a graviton is a hypothetical elementary particle mediating the force of gravity. A graviton, if it exists, must be massless as the gravitational force has unlimited range and must have a spin of 2 as gravity is a second-rank tensor field. In the brane-world context, gravitons pose serious theoretical difficulties. At high energies, i.e., processes with energies close or above the Planck scale, because of infinities arising due to quantum effects, a gravitation is non-renormalizable. In string theory, gravitons, as are all other particles, are states of strings rather than point particles. Hence, the infinities do not appear, while the low-energy behavior can still be approximated by a quantum field theory of point particles. The high concentration of gravitons near the localized Planck brane leads to a natural solution to the hierarchy problem in a universe with two branes. For the particular geometry that solves Einstein's equations, when we go out some distance in an extra dimension, we see an exponentially suppressed gravitational force. This is remarkable because it means that a huge separation of mass scales—sixteen orders of magnitude—can result from a relatively modest separation of branes. If we are living on the second brane (not the Planck brane), we would find that gravity was very weak. Such a moderate distance between branes is not difficult to achieve and is many orders of magnitude smaller than that necessary for the large-extra-dimensions scenario just discussed. A localized graviton plus a second brane separated from the brane on which the standard model of particle physics is housed provides a natural solution to the hierarchy problem—the problem of why gravity is so incredibly weak. The strength of gravity depends on location, and away from the Planck brane it is exponentially suppressed. We can think of the brane geometry, in particular the IR brane as equivalent to a needed symmetry to solve a set of equations. This construction permits (assuming K is a constant picked to fit brane-world requirements):

$$S_5 = \int d^4 x \cdot \int_{-\pi}^{\pi} d\theta \cdot R \cdot \left\{ \frac{1}{2} \cdot (\partial_M \phi)^2 - \frac{m_5^2}{2} \cdot \phi^2 - K \cdot \phi \cdot [\delta(x_5) + \delta(x_5 - \pi \cdot R)] \right\}. \qquad (7)$$

Here, what is called $m_5^2$ can be linked to Kaluza Klein "excitations" (Sundrum, 2005) via (for a number $n > 0$):

$$m_n^2 \equiv \frac{n^2}{R^2} + m_5^2. \tag{8}$$

To build the Kaluza–Klein theory, one picks an invariant metric on the circle $S^1$ that is the fiber of the $U(1)$-bundle of electromagnetism. In this discussion, an *invariant metric* is simply one that is invariant under rotations of the circle. We are using a variant of that construction via Eqn. (8). Note that In 1926, Oskar Klein proposed that the fourth spatial dimension is curled up in a circle of very small radius, so that a particle moving a short distance along that axis would return to where it began. The distance a particle can travel before reaching its initial position is said to be the size of the dimension. This extra dimension is a compact set, and the phenomenon of having a space-time with compact dimensions is referred to as compactification. In modern geometry, the extra fifth dimension can be understood to be the circle group *U(1)*, as electromagnetism can essentially be formulated as a gauge theory on a fiber bundle, the circle bundle, with gauge group *U*(1). Once this geometrical interpretation is understood, it is relatively straightforward to replace *U*(1) by a general Lie group.

Now, when we are looking at an addition of a second scalar term (Beckwith, 2006b) of opposite sign, but of equal magnitude, where:

$$S_5 = -\int d^4x \cdot V_{eff}\left(R_{phys}(x)\right) \to -\int d^4x \cdot \tilde{V}_{eff}\left(R_{phys}(x)\right). \tag{9}$$

Briefly, in classical mechanics we derive equations of motion by the principle of stationary action. In contrast, in quantum mechanics, the amplitudes of all possible motions are added in a path integral. However, if the action is replaced by the effective action, we derive equations of motion for the vacuum expectation values of the fields from the requirement that the effective action be stationary. E.g., a field φ with a potential $V(\varphi)$, at a low temperature, will not settle in a local minimum of $V(\varphi)$, but in a local minimum of the effective potential, which can be read off from the effective action. This gives (Sundrum, 2005):

$$\tilde{V}_{eff}\left(R_{phys}(x)\right) = \frac{K^2}{2 \cdot m_5} \cdot \frac{1 + \exp(m_5 \cdot \pi \cdot R_{phys}(x))}{1 - \exp(m_5 \cdot \pi \cdot R_{phys}(x))} + \frac{\tilde{K}^2}{2 \cdot \tilde{m}_5} \cdot \frac{1 - \exp(\tilde{m}_5 \cdot \pi \cdot R_{phys}(x))}{1 + \exp(\tilde{m}_5 \cdot \pi \cdot R_{phys}(x))}. \tag{10}$$

This above system has a metastable vacuum for a given special value of $R_{phys}(x)$. Start with (Beckwith, 2006b):

$$\Psi \propto \exp(-\int d^3x_{space} d\tau_{Euclidian} L_E) \equiv \exp\left(-\int d^4x \cdot L_E\right) \tag{11}$$

$$L_E \geq |Q| + \frac{1}{2} \cdot \left(\tilde{\phi} - \phi_0\right)^2 \{\ \} \xrightarrow[Q \to 0]{} \frac{1}{2} \cdot \left(\tilde{\phi} - \phi_0\right)^2 \cdot \{\ \}. \tag{12}$$

Part of the integrand in Eqn. (11) is known as an action integral, $S = \int L\, dt$, where L is the Lagrangian of the system. We also assume an Euclidean time change via $\tau = i \cdot t$, which has the effect of inverting the potential to emphasize Coleman's (Coleman, 1977) quantum bounce hypothesis. In that hypothesis, $L$ is the Lagrangian with a vanishing kinetic energy contribution, i.e. $L \to V$, where $V$ is a potential whose graph is 'inverted' by the Euclidian time. This procedure permits us, if we undo $\tau = -it$, to refer to a quantum mechanical theorem about phase evolution to address Guth's (1981) question about a generic vacuum state. Here, the spatial dimension $R_{phys}(x)$ is defined so that (Beckwith, 2006b):

$$\tilde{V}_{eff}\left(R_{phys}(x)\right) \approx \text{constant} + \tfrac{1}{2} \cdot \left(R_{phys}(x) - R_{critical}\right)^2 \propto \tilde{V}_2(\tilde{\phi}) \propto \frac{1}{2} \cdot \left(\tilde{\phi} - \phi_C\right)^2, \tag{13}$$

and (Beckwith, 2006a)

$$\{\ \} = 2 \cdot \Delta \cdot E_{gap}. \tag{14}$$

We note that the quantity $\{\ \} = 2 \cdot \Delta \cdot E_{gap}$ has a shift in minimum energy values between a false vacuum minimum energy value, $E_{false\ min}$, and a true vacuum minimum energy $E_{true\ min}$, with the difference in energy reflected in Eqn. (14). This false vacuum concept, initially promoted by Sidney Coleman in 1977, has been used, in varying degrees by cosmologists and condensed matter theorists for decades.

## USING OUR BOUND TO THE COSMOLOGICAL CONSTANT

We use our bound to the cosmological constant to obtain a conditional escape of gravitons from an early universe brane. To begin, we present conditions (Leach and Lesame, 2005) for gravitation production. Here $R$ is proportional to the scale factor 'distance':

$$B^2(R) = \frac{f_k(R)}{R^2}. \tag{15}$$

Also, there exists an 'impact parameter':

$$b^2 = \frac{E^2}{P^2}. \tag{16}$$

This leads, practically, to a condition of 'accessibility' via $R$ so defined with respect to 'bulk dimensions':

$$b \geq B(R) \tag{17}$$

$$f_k(R) = k + \frac{R^2}{l^2} - \frac{\mu}{R^2}. \tag{18}$$

Here, k = 0 for flat space, k = -1 for hyperbolic three space, and k = 1 for a three sphere, while a radius of curvature:

$$l \equiv \sqrt{\frac{-6}{\Lambda_{5-dim}}}. \tag{19}$$

This assumes a negative bulk cosmological constant $\Lambda_{5-dim}$ and that $\mu$ is a five-dimensional Schwartzshield mass. We assume emission of a graviton from a bulk horizon via scale factor, so $R_b(t) = a(t)$. Then we have a maximum effective potential of gravitons defined via:

$$B^2(R_t) = \frac{1}{l^2} + \frac{1}{4 \cdot \mu}. \tag{20}$$

This leads to a bound with respect to release of a graviton from an anti De Sitter brane (Leach and Lesame, 2006) as:

$$b \geq B(R_t). \tag{21}$$

In the language of general relativity, anti de Sitter space is the maximally symmetric vacuum solution of Einstein's field equation with a negative cosmological constant Λ. Mathematically, anti de Sitter space can be:

$$AdS_n \equiv \frac{SO(2, n-1)}{SO(1, n-1)}. \tag{22}$$

This formulation gives to $AdS_n$ an homogeneous space structure related to branes. Our universe is a five-dimensional anti de Sitter space and the elementary particles, except for the graviton, are localized on a (3 + 1)-dimensional brane. In this setting, branes, and *p*-branes, are spatially extended objects that appear in string theory. The variable *p* refers to the dimension of the brane; a 0-brane is a zero-dimensional particle, a 1-brane is a string, a 2-brane is a "membrane," etc. Each *p*-brane sweeps out a (*p*+1)-dimensional world-volume as it propagates through spacetime.

To link di quark contributions to a cosmological constant, we make four claims (Beckwith, 2006b):

Claim 1: It is possible to redefine $l \equiv \sqrt{-6/\Lambda_{5-dim}}$ as:

$$l_{eff} = \sqrt{\left|\frac{6}{\Lambda_{eff}}\right|}. \tag{23}$$

Proof of Claim 1: There is a way, for finite temperatures for defining a given four-dimensional cosmological constant; via a Park, Kim, and Tamarayan (2002) formulation, if we define:

$$k^* = \left(\frac{1}{\text{'AdS curvature}}\right). \tag{24}$$

Park, Kim, and Tamarayan (2002) note that if we have a 'horizon' temperature term:

$$U_T \propto (external\ \ temperature), \tag{25}$$

we can define a quantity:

$$\varepsilon^* = \frac{U_T^4}{k^*}. \tag{26}$$

Then there exists a relationship between a four-dimensional version of the $\Lambda_{eff}$, which may be defined by noting:

$$\Lambda_{5-dim} \equiv -3 \cdot \Lambda_{4-dim} \cdot \left(\frac{U_T}{k^{*3}}\right)^{-1} \propto -3 \cdot \Lambda_{4-dim} \cdot \left(\frac{external\ \ temperature}{k^{*3}}\right)^{-1}, \tag{27}$$

so:

$$\Lambda_{5-dim} \xrightarrow[external\ \ temperature \to small]{} \text{large value}, \tag{28}$$

and set:

$$\left|\Lambda_{5-dim}\right| = \Lambda_{eff}. \tag{29}$$

In working with these values, we should pay attention to how $\cdot\Lambda_{4-dim}$ is defined by Park, et al.

$$\cdot\Lambda_{4-dim} = 8 \cdot M_5^3 \cdot k^* \cdot \varepsilon^* \xrightarrow[external\ \ temperature \to 3\ Kelvin]{} (.0004eV)^4 \tag{30}$$

Here, we define $\Lambda_{eff}$ as being an input from Eqn. (4) to Eqn. (5) to Eqn. (6) partially due to:

$$\Delta\Lambda_{total}\Big|_{effective} = \lambda_{other} + \Delta V$$
$$\xrightarrow[\Delta V \to end\ chaotic\ inflation\ potential]{} \Lambda_{observed} \cong \Lambda_{4-dim}(3\ Kelvin). \tag{31}$$

This, for potential $V_{min}$, is defined via transition between the first and the second potentials of Eqn. (5) and Eqn. (6):

$$B_{eff}^2(R_t) = \frac{1}{l_{eff}^2} + \frac{1}{4 \cdot \mu}.  \quad (32)$$

Claim 2: $R_b(t) = a(t)$ ceases to be definable for times where the upper bound to the time limit is in terms of Planck time and in fact the entire idea of a de Sitter metric is not definable in such a physical regime (Beckwith, 2006b). This is a given in standard inflationary cosmology where traditionally the scale factor in cosmology is a parameter of the Friedmann-Lemaître-Robertson-Walker model and a function of time, which represents the relative expansion of the universe; it relates physical coordinates to *comoving coordinates*. For the FLRW model:

$$L = \overline{\overline{\lambda}} \cdot a(t),  \quad (33)$$

where L is the physical distance $\overline{\overline{\lambda}}$ is the distance in comoving units, and a(t) is the scale factor. While general relativity allows formulating the laws of physics using arbitrary coordinates, comoving coordinate choices are natural, easy to work with options; they assign constant spatial values to *comoving observers* who perceive the universe as isotropic. They are called comoving observers because they move with the Hubble flow. *Comoving distance*, the distance between two points measured along a path of constant cosmological time, can be computed by using $t_e$ as the lower limit of integration as a time of emission:

$$\overline{\overline{\lambda}} \equiv \int_{t_e}^{t} \frac{c \cdot dt'}{a(t')}.  \quad (34)$$

Claim 2 breaks down completely when one has a strongly curved space, which is what we would expect in the first instant of less than Planck time evolution of the nucleation of a new universe.

Claim 3: Eqn. (4) has a first potential for a di quark nucleation procedure, just before a defined Planck's time $t_P$. However, the cosmological constant prior to time $t_P$ was likely far higher, perhaps between values of the observed cosmological constant of today and the QCD tabulated cosmological constant, which is $10^{120}$ time greater. i.e.,

$$b^2 \geq B_{eff}^2(R_t) = \frac{1}{l_{eff}^2} + \frac{1}{4 \cdot \mu},  \quad (35)$$

which furthermore,

$$\left. \frac{1}{l_{eff}^2} \right|_{t \leq t_P} \gg \left. \frac{1}{l_{eff}^2} \right|_{t \equiv t_P + \Delta(time)}.  \quad (36)$$

So, then there would be a great release of gravitons at or about time $t_P$ (Beckwith, 2006b).

Claim 4: Few gravitons (Beckwith, 2006b) would be produced significantly after time $t_P$.

Proof of Claim 4 comes as a result of temperature changes after the initiation of inflation and changes in value of :

$$\left( \Delta l_{eff} \right)^{-1} = \left( \sqrt{\left| \frac{6}{\Lambda_{eff}} \right|} \right)^{-1} \propto \Delta \left( external \;\; temperature \right).  \quad (37)$$

## BRANE-WORLD AND DI QUARK LEAST ACTION INTEGRALS

Now for the question we are raising: Can we state the following for initial conditions of a nucleating universe?

$$S_5 = -\int d^4x \cdot \tilde{V}_{eff}\left(R_{phys}(x)\right) \propto \left(-\int d^3x_{space} d\tau_{Euclidian} L_E\right) \equiv \left(-\int d^4x \cdot L_E\right). \tag{38}$$

Instead, we should consider what can be done with S-S' instanton physics (Beckwith, 2006a), and the Bogolmyi inequality, in order to take into account baryogenesis, In physical cosmology, baryogenesis is the generic term for hypothetical physical processes that produced an asymmetry between baryons and anti-baryons in the very early universe, resulting in the substantial amounts of residual matter that comprise the universe today. $L_E$ is almost the same as Eqn. (12) above and requires elaboration of Eqn. (13) above. We should think of Eqn. (13) happening in the Planck brane mentioned above. Keep in mind that there are many baryogenesis theories in existence, The fundamental difference between baryogenesis theories is the description of the interactions between fundamental particles, and what we are doing with di quarks is actually one of the simpler ones.

## DI QUARK POTENTIAL SYSTEMS AND THE WHEELER DE-WITT EQUATION

The Penn State University developed (Ashtekar, Pawlowski, and Singh;s 2006a) quantum bounce gives a discrete version of the Wheeler De Witt equation, we begin:

$$\psi_\mu(\phi) \equiv \psi_\mu \cdot \exp(\alpha_\mu \cdot \phi^2), \tag{39}$$

as well as an energy term:

$$E_\mu = \sqrt{A_\mu \cdot B_\mu} \cdot m \cdot \hbar \tag{40}$$

$$\alpha_\mu = \sqrt{B_\mu / A_\mu} \cdot m \cdot \hbar. \tag{41}$$

This is for a 'cosmic' Schrodinger equation as given by:

$$\tilde{\tilde{H}} \cdot \psi_\mu(\phi) = E_\mu(\phi). \tag{42}$$

This has $V_\mu$ is the eigenvalue of a so-called volume operator. So:

$$A_\mu = \frac{4 \cdot m_{pl}}{9 \cdot l_{pl}^9} \cdot \left(V_{\mu+\mu_0}^{1/2} - V_{\mu-\mu_0}^{1/2}\right)^6, \tag{43}$$

and:

$$B_\mu = \frac{m_{pl}}{l_{pl}^3} \cdot (V_\mu). \tag{44}$$

In addition Ashtekar et al. (Ashtekar, Pawlowski, and Singh, 2006a, 2006b) work with as a simplistic structure with a revision of the differential equation assumed in Wheeler-De Witt theory to a form characterized by $\partial^2/\partial\phi^2 \cdot \Psi \equiv -\Theta \cdot \Psi$ and $\Theta \neq \Theta(\phi)$. This will lead to $\Psi$ having roughly the form alluded to in Eqn. (33), which in early universe geometry will eventually no longer be $L^P$ but will have a discrete geometry. This may permit an early universe 'quantum bounce' and an outline of an earlier universe collapsing, and then being recycled to match present day inflationary expansion parameters. The main idea behind the quantum theory of a (big) quantum bounce is that, as density approaches infinity, so the behavior of the quantum foam changes. The foam is a qualitative description of the turbulence that the phenomenon creates at extremely small distances of the order of the Planck length. At such small scales of time and space the uncertainty principle allows particles and energy to briefly come into existence, and then annihilate, without violating conservation laws. As the scale of time and space being discussed shrinks, the energy of the virtual particles increases. At sufficiently small scale space is not smooth as would be expected from observations at larger scales. This is where Ashtekar and other researchers (Ashtekar, Pawlowski, and Singh, 2006a, 2006b) put in their variation from earlier Wheeler–De Witt theory, with discrete segments of space to work with, and discrete time intervals as a lower bound to cosmology. Here $V_\mu$ is the eigenvalue of a so called volume operator and we need to keep in mid that the main point made above, is that a potential operator based upon a quadratic term leads to a Gaussian wave function with an exponential similarly dependent upon a quadratic $\phi^2$ exponent. This permits a match up with Eqn. (11) and Eqn. (12) above, with some additional curved space structure considerations thrown in.

## DETECTING GRAVITONS AS SPIN 2 OBJECTS WITH AVAILABLE TECHNOLOGY

To briefly review what we can say now about standard graviton detection schemes, as mentioned above, Rothman (Rothman, and Boughn, 2006) states that Dyson seriously doubts we will be able to detect gravitons via present detector technology. The conundrum is that if one defines the criterion for observing a graviton as:

$$\frac{f_\gamma \cdot \sigma}{4 \cdot \pi} \cdot \left(\frac{\alpha}{\alpha_g}\right)^{3/2} \cdot \frac{M_s}{R^2} \cdot \frac{1}{\varepsilon_\gamma} \geq 1 \ . \tag{45}$$

Here,

$$f_\gamma = \frac{L_\gamma}{L} \ . \tag{46}$$

This has $L_\gamma/L$ a graviton sources luminosity divided by total luminosity and $R$ as the distance from the graviton source, to a detector. Furthermore, $\alpha = e^2/\hbar$ and $\alpha_g = Gm_p^2/\hbar$ a constant while $\varepsilon_\gamma$ is the graviton P.E. A datum to consider is that the probability of graviton interaction with the detector 'matter' is of the order of 10 to the -60 power, whereas that for a corresponding photon would be significant orders of magnitude higher. As stated in the manuscript, the problem then becomes determining a cross section $\sigma$ for a graviton production process and $f_\gamma = L_\gamma/L$. Here, a four-dimensional graviton emission cross section goes like 1/M. The existence of branes is relevant to graviton production. In addition, it would permit us to give in confirmation of the existence of Ashtekar's candidate for a discrete wave functional for a modified Wheeler de Witt equation we would write up as follows. The point is that if we understand the contribution of Eqn. (45) above to space time dynamics, we will be able to confirm or falsify the existence of space-time conditions as given by a non $L^P$ structure as implied below. This will entail either confirming or falsifying the structure given to $\Theta$. Also, and more importantly the above mentioned $\Theta$ is a difference operator, allowing for a treatment of the scalar field as an 'emergent time', or 'internal time' so that one can set up a wave functional built about a Gaussian wave functional defined via:

$$\max \tilde{\Psi}(k) = \tilde{\Psi}(k)\big|_{k \equiv k^*} \ . \tag{47}$$

This is for a crucial 'momentum' value:

$$p_\phi^* = -\left(\sqrt{16 \cdot \pi \cdot G \cdot \hbar^2/3}\right) \cdot k^* \ , \tag{48}$$

and:

$$\phi^* = -\sqrt{3/16 \cdot \pi G} \cdot \ln|\mu^*| + \phi_0 \ . \tag{49}$$

Which leads to, for an initial point in 'trajectory space' given by the following relation $(\mu^*, \phi_0) =$ (initial degrees of freedom [dimensionless number] ~'eigenvalue of 'momentum', initial 'emergent time ') So that if we consider eigen functions of the De Witt (difference) operator, as contributing toward:

$$e_k^s(\mu) = (1/\sqrt{2}) \cdot [e_k(\mu) + e_k(-\mu)]. \qquad (50)$$

With each $e_k(\mu)$ an eigenfunction of $\Theta$ above, we have a potentially numerically treatable early universe wave functional data set that can be written as:

$$\Psi(\mu,\phi) = \int_{-\infty}^{\infty} dk \cdot \tilde{\Psi}(k) \cdot e_k^s(\mu) \cdot \exp[i\omega(k) \cdot \phi]. \qquad (51)$$

The existence of gravitons in itself would confirm or falsify the existence of non $L^P$ structure in the early universe. This structure was deemed (Ashtekar, Pawlowski, and Singh, 2006b) crucial to reference to a revision of this momentum operation along the lines of basis vectors $|\mu\rangle$ by:

$$\hat{p}_\iota |\mu\rangle = \frac{8 \cdot \pi \cdot \gamma \cdot l_{PL}^2}{6} \cdot \mu |\mu\rangle. \qquad (52)$$

With the advent of this re definition of momentum we are seeing what Ashtekar works with (Ashtekar, Pawlowski, and Singh, 2006a, 2006b) as a simplistic structure with a revision of the differential equation assumed in Wheeler-De Witt theory to a form characterized by:

$$\frac{\partial^2}{\partial \phi^2} \cdot \Psi \equiv - \Theta \cdot \Psi. \qquad (53)$$

$\Theta$ in this situation is such that:

$$\Theta \neq \Theta(\phi). \qquad (54)$$

This, in itself, would permit confirmation of whether a quantum bounce condition existed in early universe geometry according to prediction (Ashtekar, Pawlowski, and Singh, 2006a, 2006b). Additionally, prior to brane theory we had a too crude model. Why? When we assume that a radius of an early universe—assuming the speed of light $c \equiv 1$ is of the order of magnitude $3 \cdot (\Delta t \cong t_P)$—we face a rapidly changing volume that is heavily dependent upon a first order phase transition, as affected by a change in the degrees of freedom given by $\cdot(\Delta N(T))_P$. Without gravitons and a brane-world structure, such a model is insufficient to account for dark matter production, even fails to account for baryogenesis. It also will lead to new graviton detection equipment re-configuration well beyond the scope of falsifiable models configured along the lines of simple phase transitions (Banerjee, and Gavai, 1993) given for spatial volumes (assuming c = 1) of the form:

$$\Delta t \cong t_P \propto \frac{1}{4\pi} \cdot \sqrt{\frac{45}{\pi \cdot (\Delta N(T))_P}} \cdot \left(\frac{M_p}{T^2}\right). \qquad (55)$$

## GRAVITON SPACE PROPULSION SYSTEMS

Solving this final set of experimental requirements lets us consider an ensemble of wave functions for a common generic phase evolution. If we go back to an exponential phase behavior for Eqn. (11), and go from Euclidian time to regular time, we recover a phase evolution behavior; so we can use the quantum mechanics theorem that two or more wave functions with common phase behavior contain the same physical information. Such research facilitates practical efforts for generating gravitational waves in a quantum vacuum field. To do this would move warp drive research to become a testable datum, and also investigate the role extra dimensions may play in actual space drives. Once we understand how gravitons are produced in a natural setting, it would permit us to investigate the role gravitons play in exotic propulsion systems, i.e., in the early universe model. If that does not work, we at least will know and investigate another approach using near field, macro-quantum coherent states of virtual quanta not beams of on-mass-shell quantum ejected out the "nozzle" like any rocket propellant be it atoms, photons or gravitons.

## CONCLUSION

To answer Guth (2006), when there are $10^{1000}$ vacuum states produced by String theory and when inflation produces overwhelmingly one preferred type of vacuum states over other possible vacuum states, Eqn. (55) is too inexact. Instead, we have baryogenesis consistent with Eqn. (1) for $\Delta t \approx t_P$ interval. This is, for a critical temperature $T_C$, defined in the neighborhood of an initial grid of time $\Delta t \approx t_P$. If so, baryogenesis plays a role in forming early universe wave functions that are congruent with the Wheeler De Witt equation. We need to investigate whether these wave functions are congruent with a quantum bounce. If they are, it lends credence to the supposition of an earlier universe imploding due to contraction at the point of expansion at birth of our universe (Ashtekar, Pawlowski, and Singh, 2006a). According to a brane-world interpretation, gravitons produce in great number in the $\Delta t \approx t_P$ neighborhood. This depends on the 'cosmological constant' temperature. A Randall-Sundrum effective potential (Sundrum, 2005) that embeds an earlier than axion potential structure is a primary candidate for an initial configuration of dark energy. This structure would, by baryogenesis, be a shift to dark energy. We need to configure JDEM space observations to determine whether WIMPS are tied to the dark energy released after a $\Delta t \approx t_P$ interval.

## NOMENCLATURE

$\alpha = e^2/\hbar$

$\alpha_g = G \cdot M_P^2/\hbar$

$a(t)$ = scale factor proportional to $t^P$

$f_\gamma = \breve{L}_\gamma/\breve{L}$

$\breve{L}_\gamma = 7.9 \times 10^{14} \, ergs/s$

$\breve{L}$ = general background luminosity

$L_{pl}$ = Plank length = 10-33 cm

$\sigma$ = graviton production cross section proportional to $1/M$

$M = \left(M_P^2/V_n\right)^{1/2+n}$

$M_P$ = Planck mass = $10^{19}$ GeV

$\hat{V}_n$ = early universe extra dimension 'square' volume